\documentclass[
reprint,
superscriptaddress,
 amsmath,amssymb,
 aps,
prb,
longbibliography,
]{revtex4-2}

\usepackage{graphicx}% Include figure files
\usepackage{dcolumn}% Align table columns on decimal point
\usepackage{bm}% bold math
\usepackage[colorlinks,bookmarks=false,citecolor=blue,linkcolor=green,urlcolor=blue]{hyperref}% add hypertext capabilities
\usepackage{color}

\begin{document}

\title{Scaling properties of a spatial one-particle density-matrix entropy in many-body localized systems}
\author{Miroslav Hopjan}
\author{Fabian Heidrich-Meisner}
\affiliation{Institut f\"ur  Theoretische Physik, Georg-August-Universit\"at G\"ottingen, Friedrich-Hund-Platz 1, 37077 G\"ottingen, Germany}
\author{Vincenzo Alba}
\affiliation{Institute for Theoretical Physics, Universiteit van Amsterdam, Science Park 904, Postbus 94485, 1098 XH Amsterdam, The Netherlands}

\begin{abstract}
We investigate a spatial subsystem entropy extracted from the one-particle density matrix (OPDM) of
one-dimensional disordered interacting fermions that host a many-body localized
(MBL) phase. Deep in the putative MBL regime, this OPDM entropy 
exhibits the salient scaling features of localization, 
even though it provides only an upper bound to the von-Neumann entropy.
First, we numerically show that 
the OPDM entropy of the eigenstates 
obeys an area law. Second, similar to the von-Neumann entropy,  
the OPDM entropy grows logarithmically with time after a quantum quench, 
albeit with a different prefactor. 
Both these features survive at moderately large interactions and well towards the 
transition into the ergodic phase. 
We discuss prospects for calculating the OPDM entropy using approximate numerical methods 
and for its measurement in quantum-gas experiments. 
\end{abstract}

\maketitle

\section{Introduction}\label{Introduction}

Many-body localized (MBL) systems challenge the usual paradigm of 
thermalization~\cite{Altman15,Nandkishore15,Altman18,Alet18,Abanin19}. 
While it is well-established that for non-interacting particles, 
disorder leads to Anderson localization \cite{Anderson65}, 
it has been suggested that for sufficiently strong disorder, a 
localized phase survives  in the 
presence of interactions~\cite{Gornyi05,Basko06}. Despite intense 
theoretical effort  
(see recent reviews~\cite{Altman15,Nandkishore15,Altman18,Alet18,Abanin19}), 
the scenario is not fully settled. 
On the experimental side, MBL has been investigated in 
trapped ions~\cite{Smith16}, ultracold atoms \cite{Choi16,Schreiber15,Rispoli19,Lukin19,Rubio-Abadal19},
or superconducting qubits \cite{Chen17,Roushan17,Chiaro19,Guo19}. 
Experimental signatures of MBL have been observed in the quasiperiodic 
Aubry-Andr\'{e} Fermi-Hubbard model \cite{Schreiber15,Kohlert19},
the disordered Ising model \cite{Smith16}, the disordered Bose-Hubbard model (BHM) \cite{Choi16,Rubio-Abadal19},  and  
the quasiperiodic Aubry-Andr\'{e} Bose-Hubbard model \cite{Lukin19,Rispoli19}. 

Entanglement-related measures, such as the von-Neumann entropy, display 
several intriguing behaviors in the putative MBL phase. 
First, a distinctive feature of 
localization is that eigenstates exhibit 
area-law entanglement \cite{Bauer13,Kjall14,Friesdorf15}, in stark contrast 
with the volume law expected in clean systems. 
Second, the entanglement entropy grows logarithmically after global 
quenches~\cite{Znidaric08,Bardarson12,Serbyn13b,Huang_2021}, which is regarded as a 
``smoking gun'' evidence for MBL. Indeed, 
this is different in Anderson-localized systems, 
where the entanglement entropy saturates, and in clean systems,  
where  a linear behavior occurs, rigorously established for integrable models ~\cite{Calabrese_2005,Chiara_2006,fagotti-2008,Alba201703516}. 
The logarithmic growth can be explained by the existence of emergent 
local integrals of motion in the MBL 
phase~\cite{Serbyn13b,Huse14,Imbrie16a,Imbrie16b}. 
Remarkably, the logarithmic growth  of the entanglement entropy has been 
observed in cold-atom experiments~\cite{Lukin19} and systems of 
superconducting qubits~\cite{Chen17,Chiaro19}.
However, measuring entanglement is a challenging task and cannot easily be scaled up to larger systems, as it 
requires full quantum state tomography \cite{Chen17},  accessing all 
the $n$-point correlation functions \cite{Lukin19,Chiaro19}, or a high-fidelity state preparation \cite{Lukin19}. 

Here, we show that a suitably defined spatial-subsystem entropy based on the {\it one-particle density matrix} 
(OPDM) computed in eigenstates and its out-of-equilibrium dynamics after a quantum 
quench contains salient information about MBL phases, akin to the behavior of the spatial entanglement entropy. 
The main motivation for studying the OPDM is that in the MBL phase, 
the eigenstates of the OPDM are localized in 
real space~ but delocalized in the ergodic phase \cite{Bera15,Bera17}. Moreover, its eigenvalues 
indicate Fock-space localization in the MBL regime \cite{Bera15,Bera17}, a defining feature of MBL
\cite{Basko06,Luitz14,Roy19,Logan19}. This is reflected in 
the OPDM being close to that of a free-fermion 
system~\cite{Bera15,Bera17,Lezama17,Sheng-Hsuan18,Buijsman18,Villalonga19,Chen20}, and its eigenmodes being a proxy for the localized quasiparticles \cite{Bera17}.
 
We focus on the OPDM restricted to a subsystem $A$ and on the associated 
entropy. For non-interacting fermionic systems, 
this coincides with the von-Neumann entropy~\cite{Peschel09,Latorre09}. 
We consider a generic model of disordered spinless fermions with nearest-neighbor 
interactions. Numerically, we show that in the MBL phase, the disorder-averaged
OPDM entropy exhibits an area law in eigenstates, similar to 
the von-Neumann entropy. This is remarkable because in the presence of interactions, 
the OPDM entropy is not a proper (spatial) entanglement measure. 

Crucially, after a quantum quench in the MBL phase, 
the OPDM entropy increases logarithmically with time, similar to the 
von-Neumann entropy. The prefactor of the logarithmic growth is non-universal, 
and it is different from that of the von-Neumann entropy. The logarithmic growth 
survives for moderately strong interactions and as the  disorder strength is decreased. 
In the non-interacting limit, i.e., for the Anderson insulator, 
the OPDM entropy saturates. 
Our results establish the OPDM entropy as an alternative diagnostic 
tool for the MBL phase. 
This could be relevant for both experiments and approximate theoretical approaches inspired by ab-initio methods. 
Importantly, provided that one has access to the correlation function~\cite{thomson2018time}, the computational cost of extracting the OPDM entropy from the correlation function is only polynomial. 
We also note that the OPDM diagnostic tool that we propose  
is not limited to the regime of weak 
interactions, in contrast with other one-body measures based on 
Anderson orbitals \cite{Detomasi19} or the self-consistent Hartree-Fock 
approximation~\cite{Weidinger18}.

The plan of the paper is following: 
In Sec.$~$\ref{sec:model}, we introduce a model of spinless fermions  with a nearest-neighbor interaction and
provide basic definitions. The OPDM entropy will be introduced in Sec.$~$\ref{sec:opdm}.
In Sec.$~$\ref{sec:numerical}, we provide details of the numerical simulations. The distributions of the OPDM entropy
for our model are discussed in Sec.$~$\ref{sec:opdm_dist}. We then numerically demonstrate
 that the disorder-averaged OPDM entropy satisfies an area law in Sec.$~$\ref{sec:area-law}. Finally, we show in Sec.$~$\ref{sec:log}
that the disorder-averaged OPDM entropy increases logarithmically in time in global quenches from product states.
We conclude in Sec.$~$\ref{sec:concl}.

\section{Model and Definitions}
\label{sec:model}

In this paper, we consider spinless fermions with a nearest-neighbor interaction
and with diagonal disorder described by the Hamiltonian 
\begin{multline}
\label{eq:ham}
H= \sum_{i=1}^L
\Big[-\frac{J}{2}(c_{i}^{\dagger}c_{i+1}^{}+h.c)\\ + V (n_{i}^{}-1/2)(n_{i+1}^{}-1/2)
+\epsilon_{i}(n_{i}^{}-1/2)\Big],
\end{multline}
where $c_{i}^{(\dagger)}$ is a fermionic creation/annihilation operator 
and $n_{i}=c_{i}^{\dagger}c_{i}$ is the fermionic density at site $i$. $L$ is 
the system size, $J$ is the hopping matrix element, $V$ is 
the strength of the nearest-neighbor 
interactions, and $\epsilon_{i}$ is a random potential drawn from a
uniform box distribution $[-W,W]$. Using a Jordan-Wigner transformation, 
Eq.~\eqref{eq:ham} can be mapped onto
a spin-$1/2$ XXZ chain with random local magnetic fields. 
For $V/J=1$, one obtains the isotropic Heisenberg model which is
a standard system where MBL physics has been 
investigated~\cite{Luitz14,Abanin19}. Here, we  
consider $V/J=1$ and $V/J=0.1$ as representative of the strong 
and weak interactions regime, respectively.
 
We will compare the behavior of the OPDM entropy to that  of  the von-Neumann entanglement entropy $S_{\rm vN}(A)$ of a subystem $A$. 
First, we split the system into two parts, $A$ and its complement $\bar A$.  
We always consider the case in which $A$ and $\bar A$ are equal to the half chain. 
Any state of the full system  $|\psi\rangle$ can be Schmidt-decomposed 
as  
\begin{equation}
\label{eq:s-decomp}
|\psi\rangle=\sum_{\mu}\sqrt{\lambda_{\mu}}
|\phi_{\mu}\rangle_{A}|\varphi_{\mu}\rangle_{\bar A}\,,
\end{equation}
where the $\sqrt{\lambda_{\mu}}$ are the Schmidt coefficients 
and $\{|\phi_{\mu}\rangle_{A}\}$ and $\{|\varphi_{\mu}\rangle_{\bar A}\}$ 
are  orthonormal bases for $A$ and $\bar A$. The von-Neumann 
entanglement entropy is given by
\begin{equation}
\label{eq:vn}
S^{}_{\rm vN}(A) =-\sum_{\mu} \lambda_{\mu}\ln \lambda_{\mu}.
\end{equation}
For a pure state $|\psi\rangle$, Eq.~\eqref{eq:s-decomp} implies that 
$S_{\mathrm{vN}}(A)=S_{\mathrm{vN}}(\bar A)$. 

\section{OPDM entropy}
\label{sec:opdm}

Our main interest is in the properties of an entropy extracted from the  one-particle density matrix (OPDM): We restrict the OPDM
$\rho^{(1)}_{ij}=\langle\psi|c_{i}^{\dagger}c_{j}|\psi\rangle$ ($1\leq i,j\leq L$) to a subsystem $A$,  
which yields   
\begin{equation}
	\label{eq:cA}
	C_{ij}^{(A)}=\langle\psi|c_{i}^{\dagger}c_{j}|\psi\rangle, ~~~i,j \in A\,,
\end{equation}
where $|\psi\rangle$ is a many-body state. $C^{(A)}$ is usually called correlation matrix.
Given the eigenvalues $n_\alpha$ of $C^{(A)}$, we define the OPDM entropy as 
\begin{equation}
\label{eq:opdm-ent}
S_{\rm OPDM}(A)=-\sum_{\alpha}\bigl(n_{\alpha}\ln(n_{\alpha})
+(1-n_{\alpha})\ln(1-n_{\alpha})\bigr).
\end{equation}
Even though we restricted the OPDM to a subsystem, we use the name OPDM entropy for simplicity.
The OPDM entropy defined here should not be confused with the entanglement of one particle
with all other ones \cite{Bera15,Bera17,DeTomasi2017}. 
For non-interacting fermions, $S_{\rm OPDM}$ coincides with the 
von-Neumann entropy~\cite{Peschel09,Latorre09}  because the reduced density matrix of system $A$ is a Gaussian 
operator, which is fully characterized by the correlation matrix $C^{(A)}$. 
Indeed, the entanglement entropy of a generic Gaussian state with 
correlation matrix $C^{(A)}$ is given by Eq.$~$\eqref{eq:opdm-ent}. 

Several remarks are in order. First, it is instructive to consider the case where $A$ is the full system. 
Clearly, in this case, $S_{\rm vN} =0$ holds. For a free fermion system, the eigenvalues $n_\alpha$ of the OPDM are the fermionic 
occupations of the single-particle orbitals and $n_\alpha=0,1$, 
with $\sum_\alpha n_\alpha=N$, where $N$ the total number 
of fermions. By using~\eqref{eq:opdm-ent}, this 
implies that the full-system OPDM entropy is zero. 
In the presence of 
interactions, this is not the case for the OPDM entropy. Specifically, upon switching on 
interactions, yet still in the MBL regime, the eigenvalues of the  
OPDM  exhibit a bimodal distribution with $n_\alpha\approx 0,1$, signaling 
that the true eigenmodes are quasiparticles. Therefore,  the full-system 
OPDM entropy of an interacting system is non-zero except in trivial limiting cases.

While the previous argument holds for $A$ being the full system, in fact, $S_{\rm OPDM}$ always
upper-bounds $S_{\rm vN}$.
This can be seen from considering a generic fermionic state and an arbitrary partitioning. Indeed, it has 
	been shown that given the set of fermionic states with a fixed 
	matrix $C^{(A)}$ [cf.~\eqref{eq:cA}], the Gaussian states maximize the 
	von-Neumann entropy~\cite{greplova-2018}. (A similar result holds 
	for bosonic states~\cite{holevo-1999,wolf-2006}. The proof relies 
	only on the strong subadditivity of the von-Neumann entropy, on 
	the invariance under local unitary operations, and its additivity 
	for tensor-product density matrices.) 
	This implies that generic fermionic states, such as eigenstates of interacting many-body systems, 
	have lower values of $S_{\rm vN}$ than the corresponding Gaussian state with the same correlation matrix $C^{(A)}$. 
	The OPDM approximation, 
	Eq.$~$\eqref{eq:opdm-ent}, applied to an arbitrary state of an interacting system, 
	projects  onto the von-Neumann entropy  of the corresponding Gaussian state, 
	which is then necessarily larger that the true $S_{\rm vN}$.	
	This holds true both for eigenstates and the out-of-equilibrium 
	dynamics, meaning that at any time, one has $S_{\rm OPDM}(t)\ge S_{\rm vN}(t)$. 	

In the following sections, we show that the OPDM entropy exhibits two 
of the hallmark features of 
MBL, namely the area-law behavior in  excited states and the logarithmic growth 
after a global quantum quench.	

\section{Numerical Method for Obtaining Eigenstates}
\label{sec:numerical}
We use exact diagonalization to compute all  eigenstates of~\eqref{eq:ham} up
to $L =18$. We consider a system with periodic boundary conditions, and we
restrict ourselves to a fixed number of fermions $N/L=1/2$, which corresponds to
zero magnetization in the spin language. We average the OPDM entropy  over $10^4$
disorder realizations for $L\leq16$ and $10^3$ disorder realizations for $L=18$.
We focus on entanglement properties of mid-spectrum eigenstates.
Precisely, for each disorder realization,
we consider eigenstates with an energy such that
$\epsilon=(E-E_\mathrm{min})/(E_\mathrm{max}-E_\mathrm{min})\approx 1/2$,
with $E_\mathrm{min}$ the ground-state energy and $E_\mathrm{max}$
the energy of the most excited state.
We use the shift-and-invert method \cite{Pietracaprina18} to target the
desired energy window. Typically, for each disorder realization, we
consider $6$ eigenstates.

\section{Distribution of $S_{\rm OPDM}$}
\label{sec:opdm_dist}

 Throughout the paper, we consider the bipartition where $A$ and $\bar{A}$ have the same length of $L/2$.
For such a partitioning we always observe that both $S_{\rm OPDM}$ and $S_{\rm vN}$ lie within the interval
$[0,L/2\log(2)]$. Moreover, for maximally entangled states, we observe $S_{\rm OPDM}= S_{\rm vN}=L/2\ln(2)$
whereas for product states (that have no entanglement),  $S_{\rm OPDM}= S_{\rm vN}=0$ (see the discussion below).
An important observation is that in the presence of interactions and disordered, $S_\mathrm{OPDM}(A)\ne S_\mathrm{OPDM}(\bar A)$.
This asymmetry is induced by disorder, yet for the disorder average, $\bar{S}_\mathrm{OPDM}(A)\approx \bar{S}_\mathrm{OPDM}(\bar A)$.

\begin{figure}[t!]
\begin{center}
\includegraphics[width=8cm,angle=0]{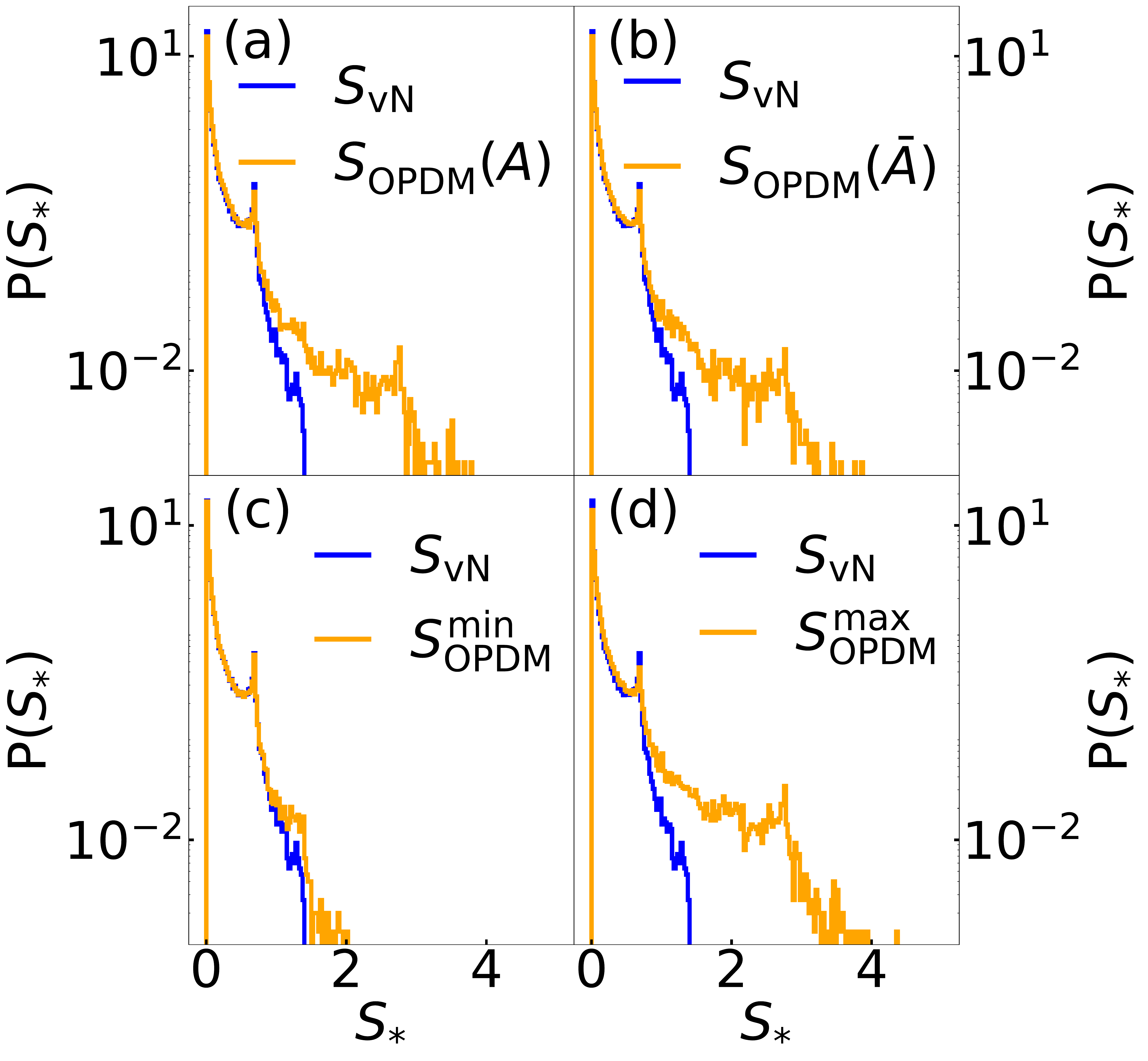}
\caption{(a), (b) Eigenstate distribution $P$ of $S_\mathrm{OPDM}(A)$ and $S_\mathrm{OPDM}(\bar A)$, respectively.
The distribution  $P(S_{\rm vN})$ (dark-shaded area) is also included. 
(c), (d) Eigenstate distribution of $S_\mathrm{OPDM}^{\rm min}$ and $S_\mathrm{OPDM}^{\rm max}$, respectively. 
 Data are averaged over $10^4$ disorder realization 
 and are obtained from $6\cdot 10^4$ eigenstates. 
 Results are for fixed $\epsilon=1, V/J=1, W/J=15$, 
 and system size $L=16$.
}\label{fig:dist}
\end{center}
\vspace{-0.8cm}
\end{figure}

Next, we study the full distribution of $S_\mathrm{OPDM}(A)$ and $S_\mathrm{OPDM}(\bar A)$, shown in Figs.~\ref{fig:dist}(a) and (b),
where the distribution of $S_{\rm vN}$ is also included.
At small values of $S_{\rm OPDM}$,  $P(S_\mathrm{OPDM}(A))\approx P(S_{\rm vN})$ 
while both $P(S_\mathrm{OPDM}(A))$ and $P(S_\mathrm{OPDM}(\bar A))$ exhibit  significant tails beyond the largest values of $S_{\rm vN}$.
 For this reason, we introduce $S_{\rm OPDM}^{\rm min}$ and $S_{\rm OPDM}^{\rm max}$ as
\begin{eqnarray}
\label{eq:opdm-min}
S_{\rm OPDM}^{\rm min}&=&{\rm min}(S_{\rm OPDM}(A),S_{\rm OPDM}(\bar A))\,\\
S_{\rm OPDM}^{\rm max}&=&{\rm max}(S_{\rm OPDM}(A),S_{\rm OPDM}(\bar A)).
\end{eqnarray}
Figures~\ref{fig:dist}(c) and (d) show the respective typical distributions in eigenstates. Clearly, 
$P(S_{\rm OPDM}^{\rm min})$ is the closest to $P(S_{\rm vN})$ as it exhibits the smallest tails at large values.
Therefore, we expect that $S_{\rm OPDM}^{\rm min}$  is the best candidate to capture the 
scaling properties of the von-Neumann entropy, which will be substantiated by the following analysis.

\begin{figure}[t!]
\begin{center}
\includegraphics[width=8cm,angle=0]{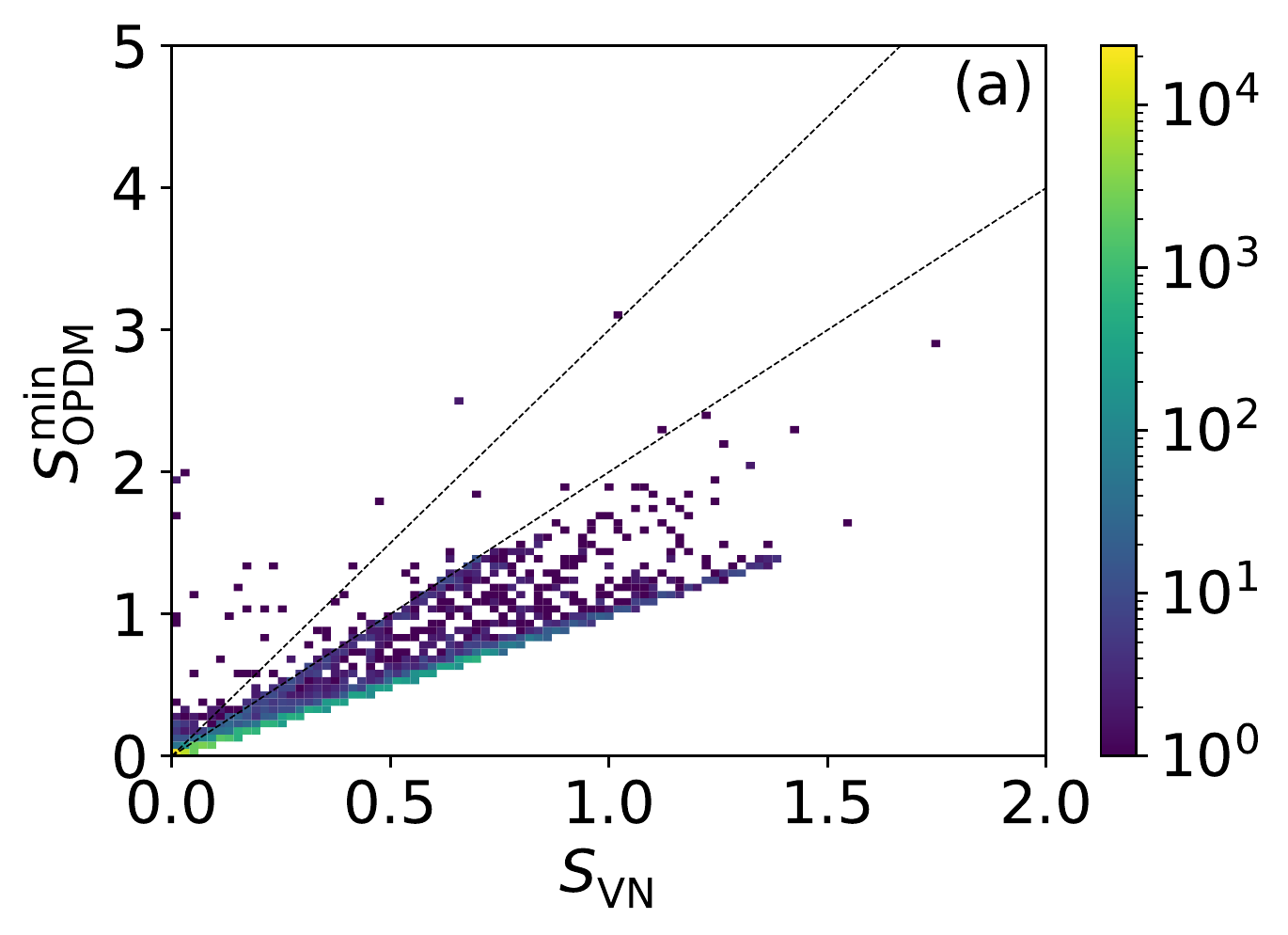}
\includegraphics[width=8cm,angle=0]{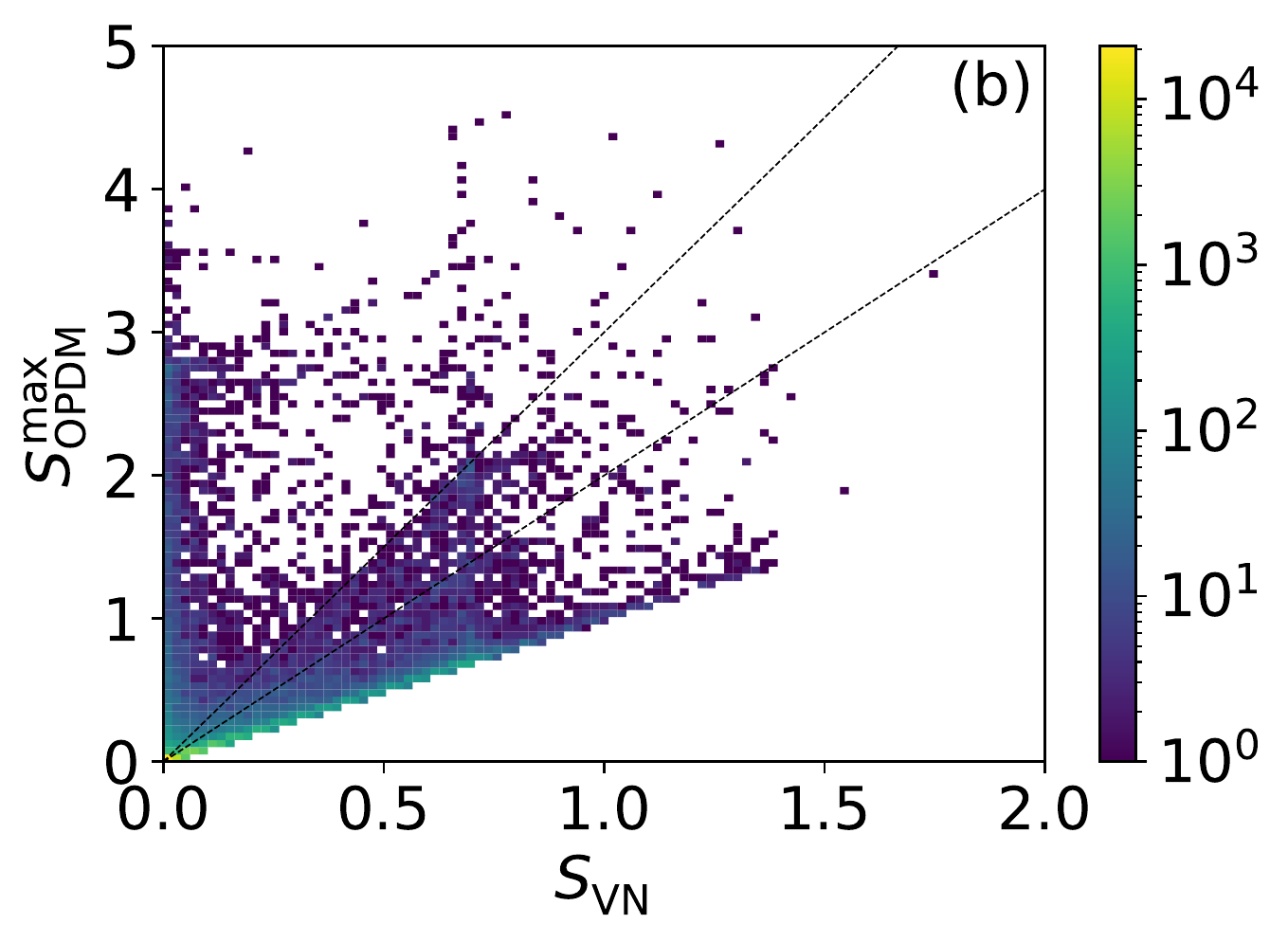}
\caption{(a) 2D histogram showing the correlation
 between von-Neumann entanglement entropy $S_{\rm vN}$ and the
 OPDM entropy $S_{\rm OPDM}^{\rm min}$, both computed in the same eigenstate.
 For each pair of $(S_\mathrm{vN},S^{\rm min}_\mathrm{OPDM})$, the color encodes the number of eigenstates with those entropies. 
 Data are averaged over $10^4$ disorder realizations 
 and are obtained from $6\cdot 10^4$ eigenstates. 
 Results are for fixed $\epsilon=1, V/J=1, W/J=15$, 
 and system size $L=16$.
 (b) The same for $S_{\rm OPDM}^{\rm max}$.
Thin dotted lines indicate $S=2S_{\rm vN}$ and $S=3S_{\rm vN}$.}
\label{correlations}
\end{center}
\vspace{-0.8cm}
\end{figure}

\section{Area law of the OPDM entropy in the MBL regime}
\label{sec:area-law}

In this section, we show that for the eigenstates of~\eqref{eq:ham}, 
the disorder-averaged OPDM entropy defined in Eq.~\eqref{eq:opdm-min} satisfies the area law. 
This behavior can be anticipated from  the limit of strong disorder, i.e., deep in the MBL 
phase. In this  limit, the 
eigenvalues of the  OPDM take the values $n_\alpha=0,1$, i.e., 
they exhibit the typical step-like behavior as for free-fermion systems.  
This signals that the MBL-localized state is close to a single 
Slater determinant~\cite{Bera15,Bera17}, for which the OPDM entropy 
coincides with the von-Neumann entropy. Since this proximity to a Slater determinant persists throughout the 
MBL regime and since the eigenmodes of the OPDM are a proxy for the localized quasiparticles \cite{Bera17},  it is natural to expect that for 
sufficiently strong disorder, the OPDM 
entropy~\eqref{eq:opdm-ent} exhibits a similar behavior as the von-Neumann entropy.

In Fig.~\ref{correlations}(a), we focus on the half-chain entanglement entropy and 
OPDM entropy for a system with $L=16$. Results are for $V/J=1$ and $W/J=15$. 
For these parameters, the system is expected to be in the MBL phase because 
the putative transition happens at 
$W_c/J\approx4$~\cite{Luitz14,Bera15,Mace20,Laflorencie20,Hopjan20}. 
Note also that for $V/J=1$, the system is far from the ``trivial'' non-interacting 
limit $V=0$.

The 2D histogram shows the correlation 
between $S_\mathrm{vN}$ (on the $x$-axis) and   $S_\mathrm{OPDM}^{\rm min}$ (on the $y$-axis) 
computed in the same eigenstate. The color scale 
denotes the number of eigenstates with a given pair of values of the entanglement entropies. 
The main conclusion from Fig.~\ref{correlations}(a) is that 
the OPDM entropy is always larger than the entanglement entropy, i.e., 
$S_{\mathrm{vN}} \le S^{\mathrm{min}}_{\rm OPDM}$ for all eigenstates. 
 This is a confirmation of the results 
of Ref.$~$\cite{greplova-2018}, yet for a many-body system. 
One can also observe that the majority of the points lies close to the 
diagonal, i.e., for most of the eigenstates, $S_{\mathrm{vN}}$ is close 
to $S^{\rm min}_{\mathrm{OPDM}}$. 
Interestingly, a  second cluster of states is visible at
$S^{\rm min}_{\mathrm{OPDM}}=2S_{\mathrm{vN}}$. This feature corresponds to resonant pairs  and is explained below. 
With increasing disorder strength, at least for fixed system size, 
all eigenstates collapse on the main diagonal 
and the minimal OPDM entropy becomes comparable to  the von-Neumann entropy. 
In the limit of strong disorder $W/J\to\infty$, the eigenstates become 
	single Slater determinants for which the equality  holds. 

For comparison, in Fig.$~$\ref{correlations}(b), we show the half-chain entanglement entropy and 
 the maximal OPDM entropy $S^{\rm max}_{\mathrm{OPDM}}$ for the same systems considered in Fig.$~$\ref{correlations}(a).
Here, one also observes  some points that are close to the 
diagonal and a  second cluster of states is visible at $S^{\rm max}_{\mathrm{OPDM}}=2S_{\mathrm{vN}}$. 
Moreover,  a third diagonal emerges along $S^{\rm max}_{\mathrm{OPDM}}=3S_{\mathrm{vN}}$
and, more importantly, there are states for which $S^{\rm max}_{\mathrm{OPDM}} \gg S_{\mathrm{vN}}\approx 0$. 
These features
render $S^{\rm min}_{\mathrm{OPDM}}$ the better object to capture the scaling properties of $S_{\rm vN}$ as compared to $S^{\rm max}_{\mathrm{OPDM}}$. 

	We now explain, using a toy two-particle system of 4 sites, that the second diagonal in Figs.$~$\ref{correlations}(a) and (b)   
	 is  due to interactions. Let us consider a two-particle state 
	\begin{equation}
	\label{state}
	|\psi\rangle=\sum_{\alpha\beta}\psi_{\alpha\beta}|\alpha,\beta\rangle=\sum_{\alpha\beta}\psi_{\alpha\beta}c^\dagger_{\alpha}c_{\beta}^\dagger|0\rangle
	\end{equation}
	 where
	$|0\rangle$ is the fermionic vacuum and the creation operator $c_{\alpha}^\dagger$ 
	creates a fermion in a localized state $\alpha$ (this can be a site, i.e., a Wannier orbital). Here, we assume that site $\alpha \in \{1,2\}$ is in subsystem 
	$A$ whereas site $\beta \in \{3,4\}$ is in $\bar A$. Such a state $|\psi\rangle$ assumes each fermion to be localized in the respective parts and
	neglects fluctuations of the fermions across the boundary (for states accounting for the fluctuations, see Appendix$~$\ref{app-2}).
	We note that the particular choice for $|\psi\rangle$ with all $\psi_{\alpha\beta}=1/2$
	was considered in Ref.~\cite{Serbyn13b} as a toy model for localized particles in an initial product state to understand the out-of-equilibrium 
	dynamics of the von-Neumann entropy.  There, it was shown that effective interactions between localized particles induce a growth of the von-Neumann entropy in time.
	
	 In those states that are described by Eq.~\eqref{state}, any (effective) interaction gives rise to correlations between the fermions
	 resulting in non-trivial eigenstates. A  nontrivial eigenstate that can be generated is a state close to the
	  superposition $|\psi'\rangle= \psi_{13}|1,3\rangle + \psi_{24}|2,4\rangle$. 
	%. 
	It is straightforward to 
	check that for $|\psi'\rangle$, the correlation matrix $C^{(A)}$ and the reduced density matrix $\rho^{(A)}$ for subsystem
	$A$ coincide and are diagonal as 
	\begin{equation}\label{density}
		\rho^{(A)}=C^{(A)}=\left(
			\begin{array}{cc}
				|\psi_{13}|^2  & 0\\
				0 & 1-|\psi_{13}|^2
			\end{array}
		\right)
	\end{equation}
	where we used the normalization condition $|\psi_{24}|^2=1-|\psi_{13}|^2$.
	For our two-particle system, the same applies to $C^{(\bar A)}$ and  $\rho^{(\bar A)}$ which 
	implies that $S^{A}_{\mathrm{OPDM}}=S^{\bar A}_{\mathrm{OPDM}}$ and thus 
	$S^{\rm min}_{\mathrm{OPDM}}=S^{\rm max}_{\mathrm{OPDM}}$. It is thus
	sufficient to consider one number $S^{}_{\mathrm{OPDM}}$ only.
	By varying the coefficient  $\psi_{13}$ one obtains  
	$0\le S_{\rm vN}\le\ln(2)$. It is also straightforward to 
	check that for any $\psi_{13}$,  $S_{\rm OPDM}=2S_{\rm vN}$. 
	
Several remarks are in order. First, the reasoning laid out  above 
	is not expected to capture the full entanglement patterns 
	in Figs.$~$\ref{correlations}(a) and (b).
	For instance, in general, a third 
	diagonal with eigenstates with $S_{\rm OPDM}=3 S_{\rm vN}$ can 
	appear, see Fig.$~$\ref{correlations}(b). This requires taking into account more complicated 
	correlations involving more than two fermions. Still,  
	 the clustering of the eigenstates around the main and 
	the second diagonal suggests that the entanglement structure is 
	 dominated by correlations involving two-body resonances across the boundary between  
	the two subsystems.

\begin{figure}[t!]
\begin{center}
\includegraphics[width=8cm,angle=0]{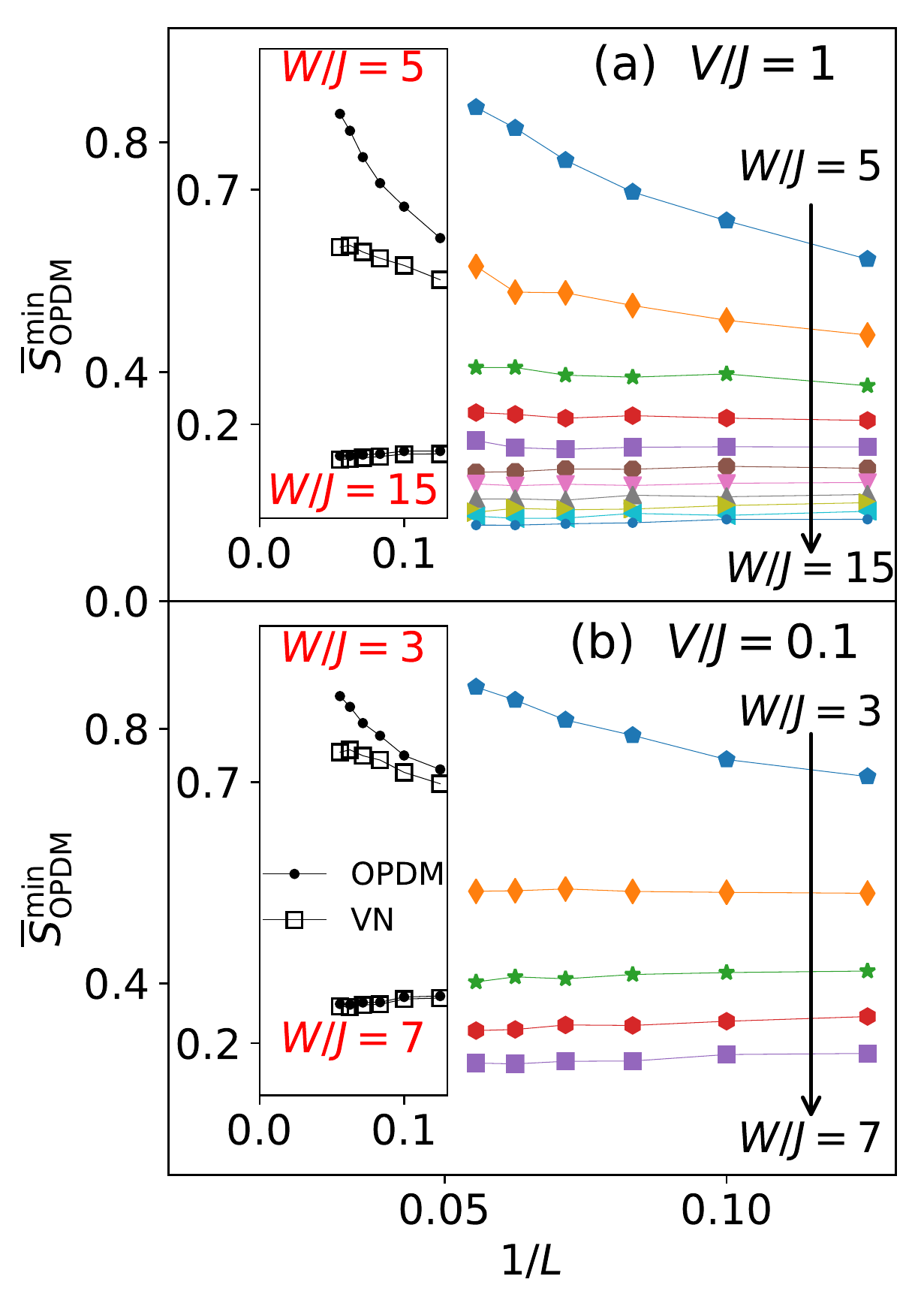}
\caption{{(a): Main panel: Disorder average of the 
 OPDM entropy $\overline{S}_{\rm OPDM}^{\rm min}$ plotted 
 as a function of $1/L$ for $V/J= 1$ and disorder strength 
 $W/J=5, \dots, 15$ (different 
 symbols). The arrow shows increasing disorder strength.
 Inset: Comparison between OPDM entropy $\overline{S}_{\rm OPDM}^{\rm min}$ 
 (full circles) and von-Neumann entropy $\overline{S}_{\rm vN}^{}$ 
 (open squares) for $V/J=1$ and $W/J=5, 15$. 
 (b): Same as in (a) for weak interactions 
 $V/J= 0.1$ and disorder strength $W/J=3, \dots, 7$. 
}}
\label{eigenstates}
\end{center}
\vspace{-0.8cm}
\end{figure}

We now demonstrate that the {\em{disorder-averaged}} OPDM entropy $\overline{S}_{\rm OPDM}^{\rm min}$ obeys the area law (the $L$-dependence of 
$\overline{S}_{\rm OPDM}^{\rm max}$ is discussed in Appendix \ref{sec:smax}). 
In Fig.$~$\ref{eigenstates}, we present the average OPDM 
entropy for the half chain as a function of $L$ for several values of 
$W/J$ and for $V/J=1$ and for $V/J=0.1$.
For $V/J=1$, standard diagnostic tools give a putative MBL transition at 
$W_c/J\approx4$ \cite{Luitz14,Bera15,Mace20,Laflorencie20,Hopjan20} (see also \cite{Devakul15,Doggen18,Chanda20,Khemani17,Suntajs19,Sierant19a,Abanin19a,Panda20,Chanda20a,Suntajs20,Sierant20a}). 
In Fig.$~$\ref{eigenstates}(a), we display the $L$-dependence of the OPDM 
entropy on the MBL side. Deep in the MBL phase (for instance, for 
$W/J\geq10$), the OPDM entropy is almost $L$-independent, implying area-law behavior
\cite{Bauer13,Kjall14,Friesdorf15}. Moreover, the OPDM entropy becomes very close to
the von-Neumann entropy upon increasing the disorder strength 
[see the inset of Fig.$~$\ref{eigenstates}(a)]. 
For the regime of weak interactions $V/J=0.1$ [see Fig.$~$\ref{eigenstates}(b)], 
smaller values of $W$ are sufficient to observe the area-law behavior. 
This is expected because upon lowering $V$, the MBL transition 
is shifted towards smaller values of 
$W$. We estimate the transition at $V=0.1J$
from standard diagnostic tools, such as the average gap ratio 
\cite{Oganesyan07,Luitz14} and 
the occupation distance measure \cite{Hopjan20}, which give $W_c/J\approx2-3$  (see Appendix ~\ref{app-1}).
An analysis of the behavior of  $\bar S_{\rm OPDM}^{\rm min}$ across the transition into the ergodic region is beyond the 
scope of the present work and left for future research.

\section{Logarithmic growth of the OPDM entropy in the MBL regime}
\label{sec:log}

Next, we discuss the time-dependence of the OPDM entropy after a global quantum quench 
deep in the MBL phase. 
We consider the evolution from initial random product states such as 
$|\psi_0\rangle = |1010\dots 1\rangle$, where $0,1$ are the initial fermionic occupations. 
We study the Hamiltonian dynamics $|\psi(t)\rangle =e^{-i {H}t}|\psi_0\rangle$ 
by using full exact diagonalization of $H$. 
For each disorder configuration, we select product states $|\psi_0\rangle$ with 
energy density $\epsilon=(\langle \psi_0| H|\psi_0\rangle-E_{\rm min})/(E_{\rm max} - 
E_{\rm min})$ that fulfills $|1/2-\epsilon| \lesssim 2\cdot 10^{-4}$, i.e., close to mid-spectrum energy density. 
We average over $200$ disorder realizations. 

In the putative MBL phase, the von-Neumann entropy grows logarithmically  after 
global quenches \cite{Bardarson12,Serbyn13b,Znidaric08,Serbyn15}, whereas 
on the ergodic side,  a ballistic or sub-ballistic entanglement growth is observed \cite{Luitz16}.  
The change of behavior happens at the eigenstate transition \cite{Serbyn15}. 
As  anticipated in Sec.~\ref{sec:opdm}, 
after a quantum quench, we observe $S_\mathrm{OPDM}(t)\geq S_\mathrm{vN}(t)$ in all states at any time. 

First, we provide a simple argument why the OPDM entropy increases logarithmically 
after a quantum quench from initial product state. 
Let us again consider a generic state in Eq.$~$(\ref{state}). 
We take a particular choice of this state, where all $\psi_{\alpha\beta}=1/2$, and 
follow the arguments of Ref.~\cite{Serbyn13b}. The initial state can be written as
$|\psi_0\rangle=\frac{1}{2}(c^\dagger_1+c^\dagger_2)(c^\dagger_3+c^\dagger_4)|0\rangle$.
For such initial state, there is no entanglement between the particle in orbitals $1,2$ and the particle in orbitals  $3,4$ at $t=0$ as can be 
verified by a direct calculation.  An effective interaction produces the time-evolved state $|\psi(t)\rangle$, which is given by
\begin{equation}
\label{state_time}
	|\psi(t)\rangle=\sum_{\alpha,\beta}\frac{1}{2}e^{-iE_{\alpha\beta}t}|\alpha,\beta\rangle\,,
\end{equation}
where the  energies $E_{\alpha\beta}$ are $E_{\alpha\beta}=\epsilon_\alpha+\epsilon_\beta+\delta E_{\alpha\beta}$
and $\epsilon_\alpha$ and $\epsilon_\beta$ are the single-particle energies, whereas $\delta E_{\alpha\beta}$ 
is  due to the interactions. It is natural to expect that 
$\delta E_{\alpha\beta}=k_{\alpha\beta}\tilde{V}e^{-x/\xi}$, with $k_{\alpha\beta}$ a 
constant, $\tilde{V}$ an effective interaction strength between the two localized particles with distance $x$ from each other, and $\xi$ the localization length, for which we  assume $x\gg\xi$. 
For the state from Eq.$~$(\ref{state_time}), the reduced density matrix $\rho^{(A)}$ coincides with the OPDM matrix $C^{(A)}$ and is given by 
\begin{equation}
	\rho^{(A)}=C^{(A)}=\frac{1}{2}\left(
\begin{array}{cc}
	1 &  F(t)/2\\
	F^*(t)/2 & 1
\end{array}
\right)\,,
\end{equation}
where $F(t)=e^{-i\Omega t}(1+e^{-i\delta \Omega t})$,
with $\Omega=\epsilon_1-\epsilon_2+\delta E_{13}-\delta E_{23}$ and 
$\delta\Omega=\delta E_{14}-\delta E_{24}+\delta E_{13}+\delta E_{23}$. 
The eigenvalues of $\rho^{(A)}$  are 
\begin{equation}
	\lambda_\pm=\frac{1}{2}\Big(1\pm\frac{|F(t)|}{2}\Big). 
\end{equation}
Note that  $F(t)$ vanishes at $t^*=\pi/\delta\Omega$ and at $t^*=\pi/2\delta\Omega$, the entanglement 
entropy $S_{\rm vN}$ has a maximum with $S_{\rm vN}=\ln(2)$. This effect is due to 
$\delta\Omega$, which reflects the presence of (effective) interactions. 
By using the definition of the OPDM entropy, one obtains that $S_{\rm OPDM}(t)=2S_{\rm vN}(t)$. 
Therefore,  at any time, the OPDM entropy differs from the von-Neumann entropy only
by a pre-factor. This pre-factor can  in general be different from 2 because the toy state
 in Eq.~\eqref{state_time} does not account for the full correlation pattern of a general many-body 
wave function. 

\begin{figure}[t!]
\begin{center}
\includegraphics[width=8cm,angle=0]{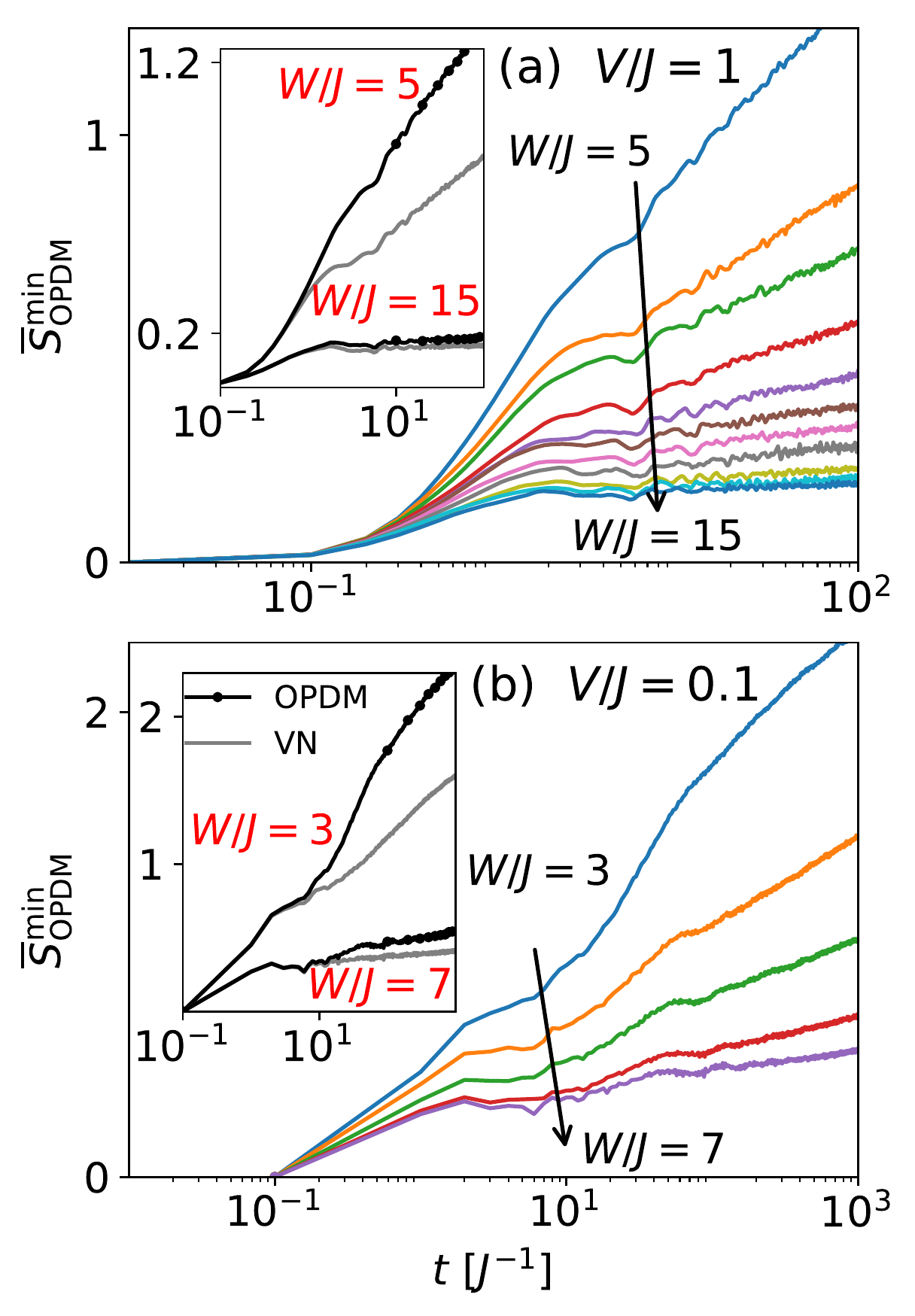}
\caption{(a): Main panel: Time evolution of the average OPDM 
 entropy $\overline{S}_{\rm OPDM}^{\rm min}$ for $L=16$, 
 $V/J= 1$ and disorder strength $W/J=5, \dots, 15$. 
 The arrow denotes  increasing disorder strength. 
 For all values of $W$, a clear logarithmic growth with a non-universal 
 prefactor is visible  at long times. 
 Inset: Comparison between the OPDM entropy $\overline{S}_{\rm OPDM}^{\rm min}$ 
 (full circles) and the von-Neumann entropy $\overline{S}_{\rm vN}^{}$ 
 (gray lines) for $L=16$, $V/J= 1$ and $W/J=5, 15$. (b) Same as in (a) for weak interactions 
 $V/J= 0.1$ and disorder strength $W/J=3, \dots, 7$. 
}
\label{time_evolution}
\end{center}
\vspace{-0.8cm}
\end{figure}
In Fig.$~$\ref{time_evolution}(a), we show the dynamics  
of the OPDM entropy $\overline{S}_{\rm OPDM}^{\rm min}$ for $V/J=1$ (strong interactions) 
and  $W/J=5, \dots, 15$ and for $V/J=0.1$ (weak interactions) and $W/J=3, \dots, 7$ 
computed numerically for $L=16$ 
(the time dependence of $\overline{S}_{\rm OPDM}^{\rm max}$ is discussed in Appendix \ref{sec:smax}).
 In both cases, the system is in the MBL phase.
For large enough times, the data exhibit a clear logarithmic increase for  
all  values of $W$. The prefactor of the logarithmic growth depends on the 
interaction strength $V$, and hence is non-universal, as for the von-Neumann entropy. 
Note that in the limit 
$W/J\to\infty$, the entropy saturates. Interestingly, the prefactors of the logarithmic 
growth of the OPDM entropy and of the von-Neumann entropy are not the same. This is 
illustrated in the insets of Fig~\ref{time_evolution}. Only in the limit 
of large $W/J$, the dynamics of the OPDM entropy becomes quantitatively the same as  the 
von-Neumann entropy. 

Finally,  interesting features appear for weak interactions 
[see Fig.~\ref{time_evolution}(b)]. First, 
longer times are needed for the logarithmic behavior to set in. 
For instance, for $V/J=0.1$, this happens for 
$t J\geq 100$. Moreover, the dynamics of the 
OPDM entropy and that of the von-Neumann entropy is the same at short times. This is highlighted 
in the insets in Fig.~\ref{time_evolution}. Clearly, the 
OPDM entropy coincides with the von-Neumann entropy up to $tJ\approx 10$. 

\section{Conclusions}
\label{sec:concl}

We provided numerical evidence that the OPDM entropy exhibits the salient 
features of the von-Neumann entropy in putative MBL phases of matter. 
Specifically, deep in MBL phases, the eigenstate OPDM entropy obeys the area law. 
Most importantly, the entropy grows logarithmically after a global quantum quench. 
Although formally, this is expected in the limit $W/J\to\infty$, we observe that 
there is a sizeable region in parameter space, i.e., interaction and disorder 
strength, where this behavior persists. 
This behavior is expected since the OPDM eigenstates approximate the localizes quasiparticles aka l-bits in the
MBL phase \cite{Bera17}.

There are several interesting directions for future work. First, our results could 
be combined with {\it ab-initio} methods for the correlation functions, e.g., 
Green's function methods \cite{BarLev14,BarLev16,Weidinger18}. This would allow 
to compute the evolution of the OPDM entropy for larger systems, and also in 
higher dimensions. Importantly, the computation of the OPDM scales only polynomial in the linear dimension. 
Moreover, it would be interesting to measure the evolution 
of the OPDM entropy in cold-atom experiments  using single-site resolution \cite{Sherson10,Lukin19,Ardila18} or in embryonic quantum computers~\cite{supremacy,smith2019simulating}.  
Finally, in contrast with the entanglement entropy, the OPDM entropy relies on 
the fermionic correlation functions, which are standard tools in condensed matter 
physics. This renders the OPDM amenable to an analytical study and, for instance, by 
using the renormalization-group techniques reviewed in Ref.~\onlinecite{Altman15}.

\begin{acknowledgements} 
We acknowledge useful discussions with J. H. Bardarson.
\end{acknowledgements}

\appendix

\section{OPDM versus von-Neumann entropy in a two-particle, four-site system}
\label{app-2}

In this section, we compare the OPDM and the von-Neumann entropy in 
a generic 2-particle fermionic state in a 4-site system with bipartition $A=\lbrace 1,2\rbrace$ and $\bar A=\lbrace 3,4\rbrace$.
Note that for such system  $S^{A}_{\mathrm{OPDM}}=S^{\bar A}_{\mathrm{OPDM}}$  and it is thus
sufficient to consider one number $S^{}_{\mathrm{OPDM}}$ only.
Particles can be thought of as localized in $A$ or as being delocalized across the boundary between $A$ and $\bar A$. With this setup, we consider a generic wavefunction  
	\begin{equation}
	\label{state_appendix}
	|\psi\rangle=\sum_{\alpha<\beta}\psi_{\alpha\beta}|\alpha,\beta\rangle=\sum_{\alpha<\beta}\psi_{\alpha\beta}c^\dagger_{\alpha}c_{\beta}^\dagger|0\rangle \,,
	\end{equation}
	 where
	$|0\rangle$ is the fermionic vacuum and the creation operator $c_{\alpha}^\dagger$ 
	creates a fermion at site $\alpha$.   
	We allow for particle fluctuations across the partition, i.e., $\alpha\in \{1,2,3\}$ and $\beta \in \{2,3,4\}$.
	The state in Eq.$~$\eqref{state_appendix} is a generalization of the state in Eq.$~$\eqref{state} discussed
	in the main text. For such state, the matrix elements of $C^{(A)}$  read
\begin{align}
	\label{eq:CA1}
	C^{(A)}_{11} &=|\psi_{12}|^2+|\psi_{13}|^2+|\psi_{14}|^2,\\
	C^{(A)}_{22} &=|\psi_{24}|^2+|\psi_{23}|^2+|\psi_{12}|^2,\\
	C^{(A)}_{12} &=\psi_{14}^*\psi_{24}+\psi_{13}^*\psi_{23},\\
	\label{eq:CA2}
	C^{(A)}_{21} &=(C^{(A)}_{12})^*,
\end{align}
and the reduced density matrix reads  
\begin{equation}
	\label{eq:rhoA}
\rho^{(A)}=\left(
	\begin{array}{cccc}
		|\psi_{12}|^2 & 0 & 0 & 0\\
		0 & C^{(A)}_{11} -|\psi_{12}|^2 & C^{(A)}_{12} & \\
		0 & C^{(A)}_{21} & C^{(A)}_{22}-|\psi_{12}|^2\\
		0 & 0 & 0 & |\psi_{34}|^2 
	\end{array}
\right) \,.
\end{equation}
From~\eqref{eq:rhoA} and~\eqref{eq:CA1}-\eqref{eq:CA2} one can construct 
the OPDM entropy and the entanglement entropy. 

\begin{figure}[t!]
\includegraphics[width=7cm]{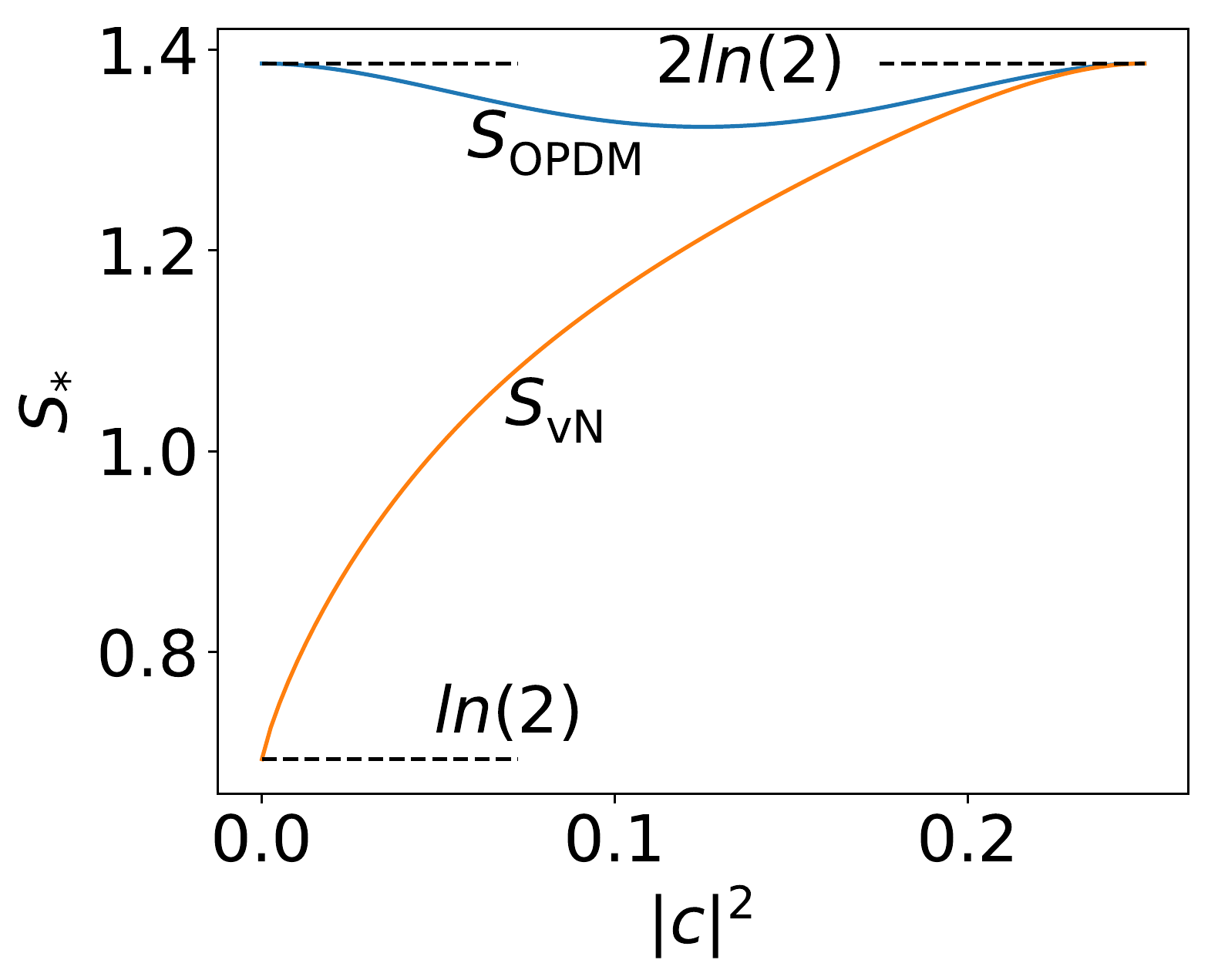}
\caption{Comparison between the von-Neumann and the 
 $\rm OPDM$ entropy for the two-fermions states in~\eqref{eq:two} as a function of
 $|c|^2$. Notice that  $S_{\rm vN}=2S_{\rm OPDM}$ for $|c|=0$ and $S_{\rm vN}=S_{\rm OPDM}$ for $|c|=1/2$. 
 The horizontal dotted lines mark the values $\ln(2)$ and $2\ln(2)$. 
}
\label{fig:comparison}
\end{figure}

Neglecting  particle fluctuations across the partition, i.e., setting $\psi_{12}=\psi_{34}=0$, 
the OPDM coincides with the reduced density matrix $\rho^{(A)}=C^{(A)}$ and is given by 
\begin{equation}\label{density_appendix}
	\rho^{(A)}=\left(
\begin{array}{cc}
	|\psi_{13}|^2+|\psi_{14}|^2 &  \psi_{14}^*\psi_{24}+\psi_{13}^*\psi_{23}\\
	\psi_{24}^*\psi_{14}+\psi_{23}^*\psi_{13} & |\psi_{24}|^2+|\psi_{23}|^2
\end{array}
\right) \,.
\end{equation}
A special case of the reduced density matrix in Eq.$~$\eqref{density_appendix} was given in the
main text in Eq.$~$\eqref{density} for which it was shown that  $S_{\rm OPDM}=2S_{\rm vN}$.

We will now discuss the clean system of $2$ fermions in $4$ sites  with translational invariance. Here, 
there are 2 relevant state. The first one is the state $|\psi'\rangle$ from Eq.$~$\eqref{state} with $|\psi_{13}|^2=|\psi_{24}|^2$ 
which is translational invariant and can be written as
\begin{equation}\label{state_appendix2}
	|\psi'\rangle=\sum_{\alpha=1}^{2}\psi_{\alpha\alpha+2}|\alpha,\alpha+2\rangle,
\end{equation}
for which we derive that $S_{\rm OPDM}=2S_{\rm vN}$ in the main text. 
The second relevant state can be written as
\begin{equation}\label{state_delocalized}
	|\psi''\rangle=\sum_{\alpha=1}^{4}\psi_{\alpha\alpha+1}|\alpha,\alpha+1\rangle,
\end{equation}
where $|\psi_{12}|^2=|\psi_{23}|^2=|\psi_{34}|^2=|\psi_{14}|^2=1/2$ and we assume the periodic boundary conditions.
Now the reduced density matrix is diagonal in the basis 
with four eigenvalues $1/4$, giving  the von-Neumann entropy $S_{\rm vN}=2\ln(2)$. At the same time, the OPDM 
matrix is diagonal with two equal eigenvalues $1/2$, which give $S_{\rm OPDM}=2\ln(2)$. This confirms our expectation
that for a maximally entangled state, the two entropies have to be equal.

The eigenstates of clean systems can be thought of as  a mixture of the states in Eq.$~$\eqref{state_appendix} and Eq.$~$\eqref{state_delocalized} 
\begin{equation}
	\label{eq:two}
	|\tilde{\psi}\rangle=\sum_{\alpha=1}^{2}\psi_{\alpha\alpha+2}|\alpha,\alpha+2\rangle+\sum_{\alpha=1}^{4}\psi_{\alpha\alpha+1}|\alpha,\alpha+1\rangle,
\end{equation}
where $|\psi_{12}|^2=|\psi_{23}|^2=|\psi_{34}|^2=|\psi_{14}|^2=:|c|^2$ and $|\psi_{13}|^2=|\psi_{24}|^2=:|c'|^2$. 
Here, due to the normalization condition, we have $|c'|^2=(1-4|c|^2)/2$. 
From~\eqref{eq:rhoA} and~\eqref{eq:CA1}-\eqref{eq:CA2}, one can 
obtain the entanglement entropy and the OPDM entropy as a function of $|c|^2$. In Fig.~\ref{fig:comparison}, we plot
 $S_{\rm vN}$ and $S_{\rm OPDM}$ as a function of $|c|^2$ assuming that both $c$ and $c'$ are real.
 As is clear from the figure,  $S_{\rm OPDM}\ge S_{\rm vN}$ for any $|c|$. 
We thus see how the entropies interpolate between states $|\psi'\rangle$ and $|\psi''\rangle$.

\section{Scaling of $\bar S_{\rm OPDM}^{\rm max}$}
\label{sec:smax}

\begin{figure}[t!]
\begin{center}
\includegraphics[width=8cm,angle=0]{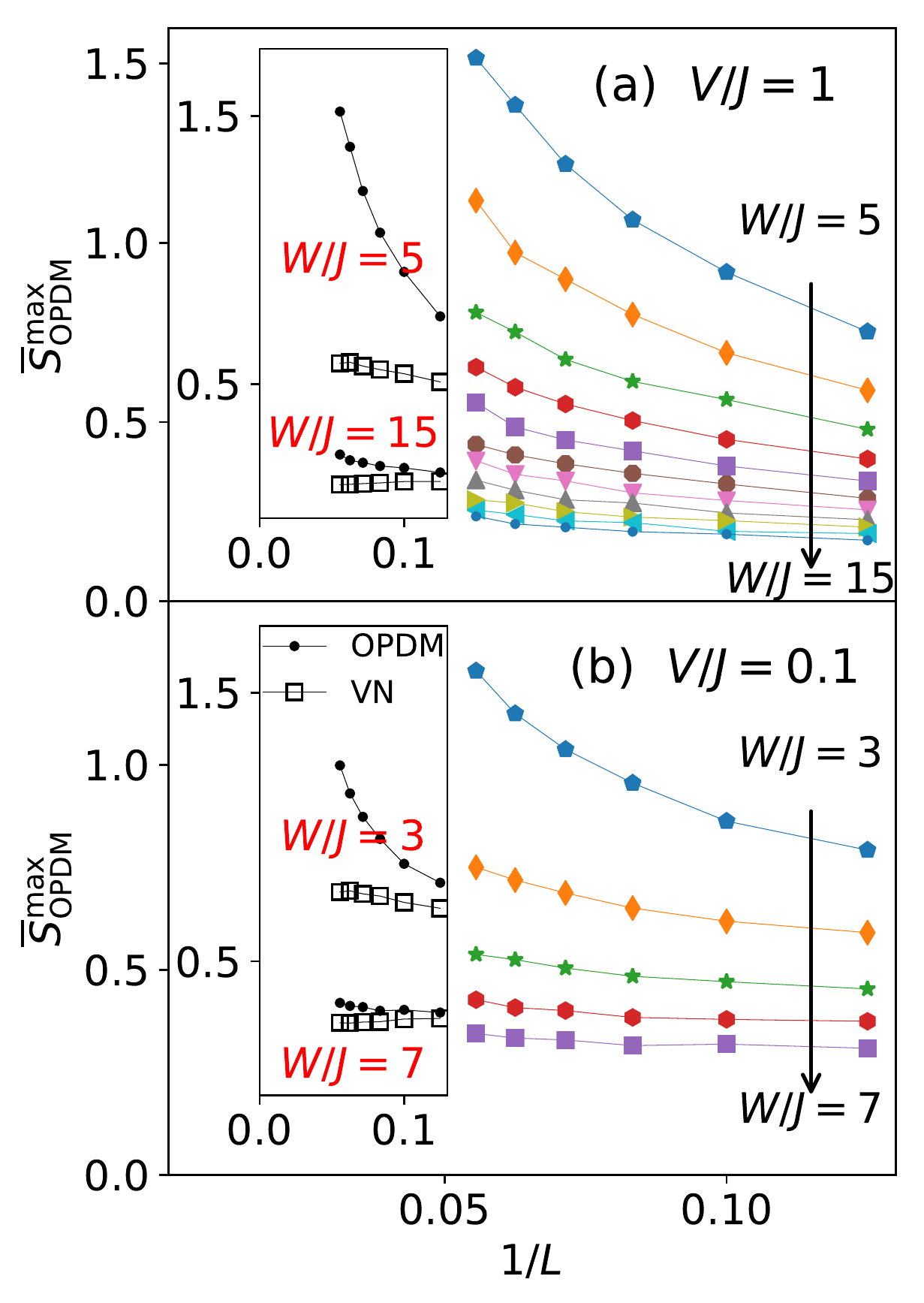}
\caption{(a): Main panel: Disorder average of the
 OPDM entropy $\bar{S}_{\rm OPDM}^{\rm max}$ plotted
 as a function of $1/L$ for $V/J= 1$ and disorder strength
 $W/J=5, \dots, 15$ (different
 symbols). The arrow shows increasing disorder strength.
 Inset: Comparison between $\bar{S}_{\rm OPDM}^{\rm max}$
 (full circles) and von-Neumann entropy $\bar{S}_{\rm vN}^{}$
 (open squares) for $V/J=1$ and $W/J=5, 15$.
 (b): Same as in (a) for weak interactions 
 $V/J= 0.1$ and disorder strength $W/J=3, \dots, 7$. 
}
\label{eigenstates2}
\end{center}
\end{figure}

\begin{figure}[t!]
\begin{center}
\includegraphics[width=8cm,angle=0]{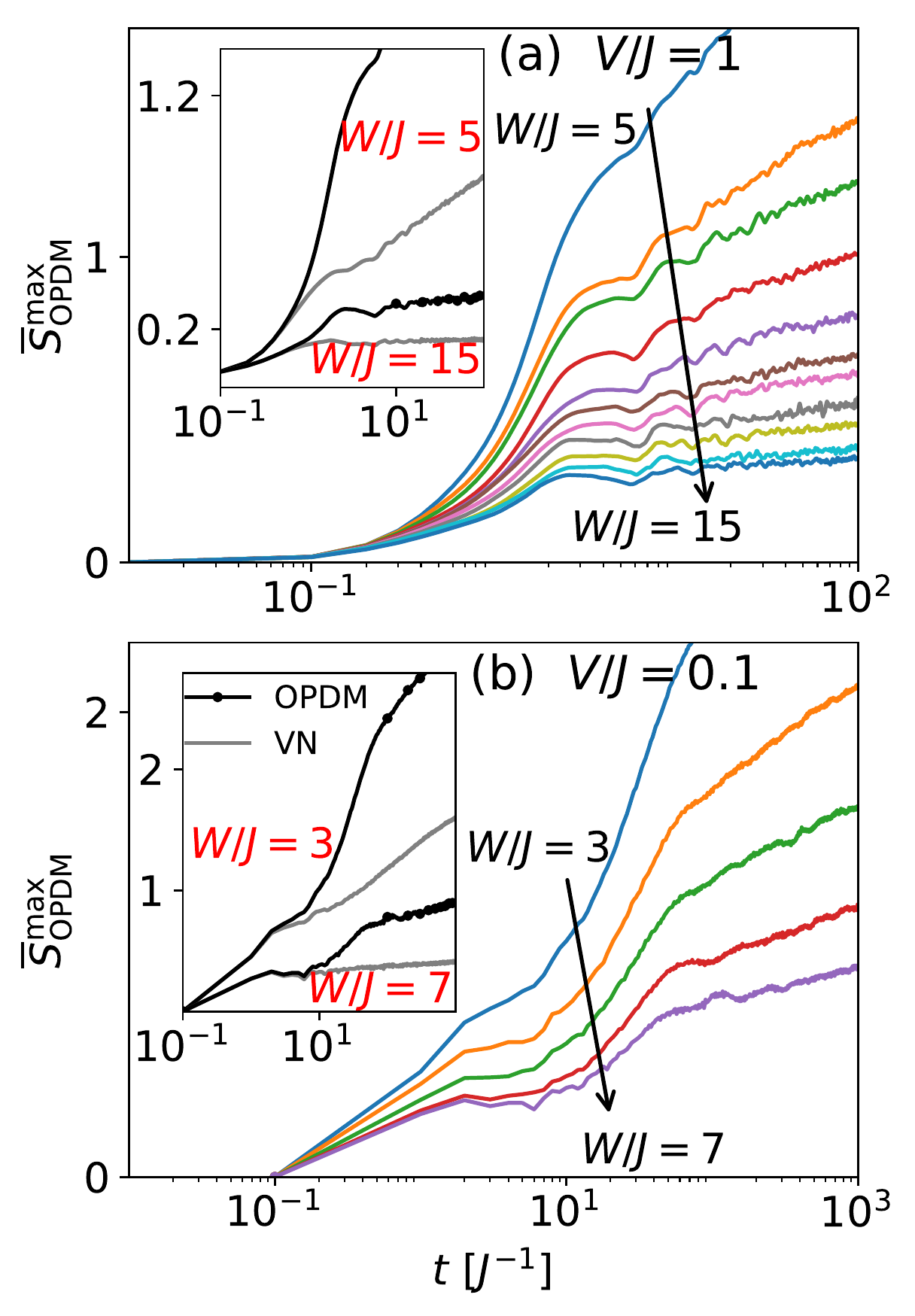}
\caption{(a): Main panel: Time evolution of the average OPDM
 entropy $\bar{S}_{\rm OPDM}^{\rm max}$ for $L=16$,
 $V/J= 1$ and disorder strength $W/J=5,\dots, 15$.
 The arrow denotes  increasing disorder strength.
 For all values of $W$, a clear logarithmic growth
 with a non-universal prefactor is visible  at long times.
 Inset: Comparison between $\bar{S}_{\rm OPDM}^{\rm max}$
 (full circles) and the von-Neumann entropy $\bar{S}_{\rm vN}^{}$
 (gray lines) for $L=16$, $V/J= 1$ and $W/J=5, 15$. Note that
 the prefactor of the logarithmic growth of the OPDM entropy and of the
 von-Neumann entropy are different. (b) Same as in (a) for weak interactions 
 $V/J= 0.1$ and disorder strength $W/J=3, \dots, 7$. }
\label{time}
\end{center}
\end{figure}

 For comparison to the behavior of $\bar S_{\rm OPDM}^{\rm min}$ shown in Fig.$~$\ref{eigenstates}, in Fig.$~$\ref{eigenstates2}, we show the average OPDM
entropy $\bar S_{\rm OPDM}^{\rm max}$ for the half chain as
a function of $L$ and for several values of
disorder strength $W/J$.
On the one hand,
 deep in the MBL phase, e.g., at $W/J\approx15$ for
$V/J=1$ [see Fig.~\ref{eigenstates2}(a)], the behavior of $\bar S_{\rm OPDM}^{\rm min}$ is  similar to
that of $\bar S_{\rm OPDM}^{\rm min}$ and  $\bar S_{\rm OPDM}^{\rm max}$. On the other hand,  $\bar{S}_{\rm OPDM}^{\rm min}$ is closer to the
von-Neumann entropy than $\bar{S}_{\rm OPDM}^{\rm max}$ as displayed in the insets of Figs.$~$\ref{eigenstates} and
\ref{eigenstates2}. This is expected since $\bar{S}_{\rm OPDM}^{\rm max}$ includes
more resonances across the two subsystems than
$\bar{S}_{\rm OPDM}^{\rm min}$.

In Fig.~\ref{time}, we show the time evolution of $\bar{S}_{\rm OPDM}^{\rm max}$
for the same set of parameters as for $\bar{S}_{\rm OPDM}^{\rm min}$  in Fig.~\ref{time_evolution}.
As  is clear from the figure, $\bar S_{\rm OPDM}^{\rm max}$ grows
logarithmically with  time after the quench. In the insets of
Fig.~\ref{time}, we compare $\bar {S}_{\rm OPDM}^{\rm max}$ and the
von-Neumann entropy. Clearly, both entropies exhibit a  logarithmic
growth but with a different non-universal prefactor. Nevertheless, as in the case of eigenstates,
we observe that $\bar{S}_{\rm OPDM}^{\rm min}$ compares better than $\bar{S}_{\rm OPDM}^{\rm max}$
 to the von-Neumann entropy as displayed in the insets of Figs.$~$\ref{time_evolution} and \ref{time}.

\section{MBL transition estimate and additional numerical data for weak interactions}
\label{app-1}

\begin{figure}[t]
\begin{center}
\includegraphics[width=7.5cm,angle=0]{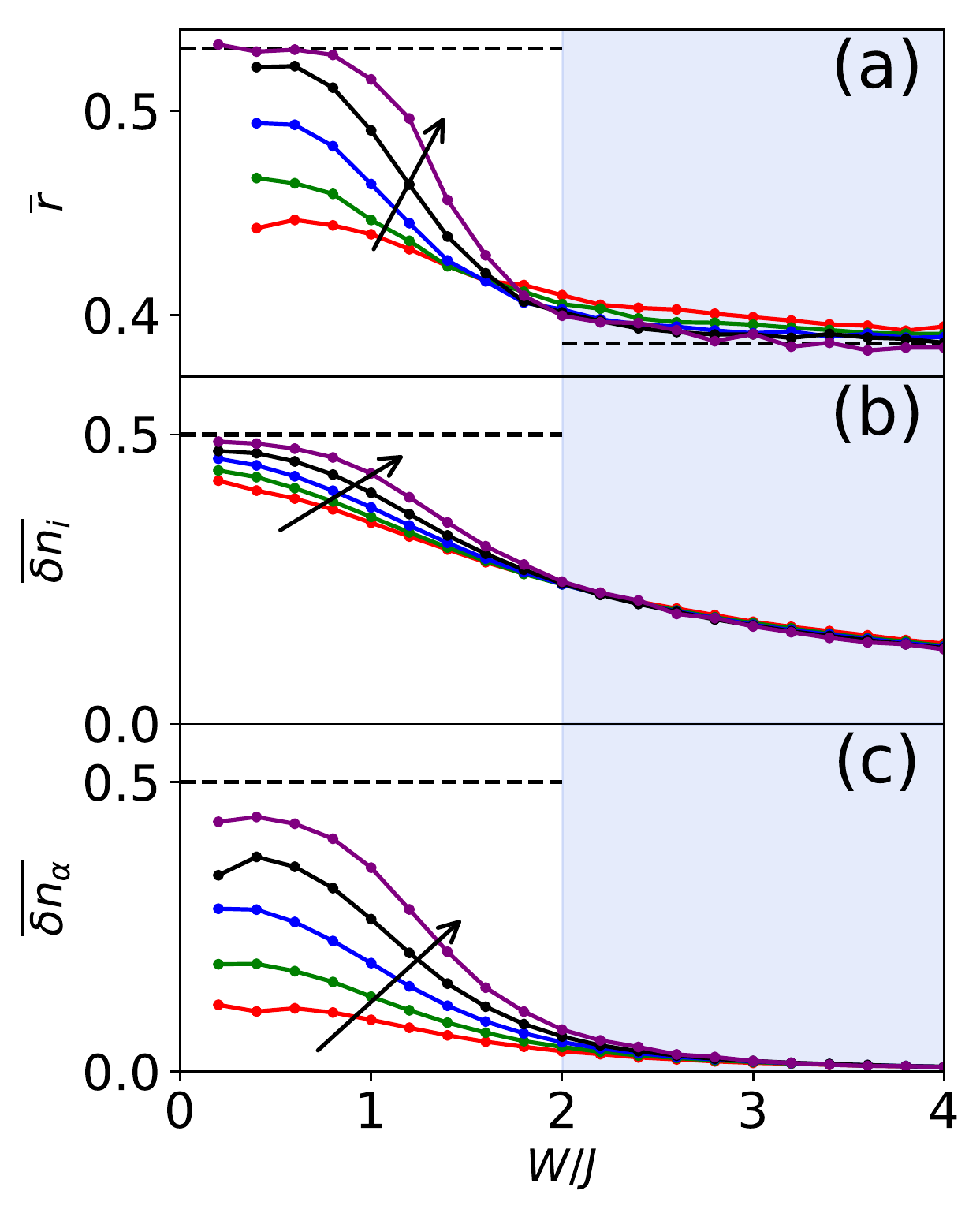}
\caption{Diagnostics of the MBL transition at $V/J=0.1$: (a) average gap ratio 
 $\bar r$, (b) average occupation distances $\overline{\delta n_{i}}$, (c) and $\overline{\delta n_{\alpha}}$.  
 $W/J$, plotted on the horizontal axis is the disorder strength. We show data  
 for $V/J=0.1$. Different curves correspond to different system 
 sizes $L=10, 12, 14, 16, 18$. Data are averaged over up to $10^5$ 
 disorder realization. The arrows denote increasing system size.
 In (a), the horizontal dashed lines denote the analytic results assuming 
 Wigner-Dyson ($\bar r\approx 0.53$) and Poisson 
 distribution ($\bar r\approx 0.38$) of the energy level spacings.
 In (b), the occupation distances are expected to attain 
 the values $\overline{\delta n_i}=1/2$ 
 in the ergodic phase (dashed lines). 
 In all panels, the shaded area is estimated to be in the 
 MBL phase. Thus, all results shown in the main text are for a disorder strength well above the transition.}
\label{transition}
\end{center}
\vspace{-0.8cm}
\end{figure}

In this section, we employ standard diagnostic tools to identify the putative 
MBL transition at  $V/J=0.1$ (see the main text). Specifically, we consider the 
average gap ratio \cite{Oganesyan07,Luitz14} and 
the occupation distance measure \cite{Hopjan20}.

We start discussing the average gap ratio $\bar r$. Given the 
eigenergies $E_n$ of the quantum many-body Hamiltonian, we first 
define the gaps $\delta_n$ as 
\begin{equation}
\delta_n\equiv E_{n+1}-E_n\,. 
\end{equation}
The gap ratio $r_n$ is defined as  
\begin{equation}
\label{equ:ratio}
0\le r_n=\mathrm{min}\{\delta_n,\delta_{n-1}\}/\mathrm{max}\{\delta_n,\delta_{n-1}\}\le1. 
\end{equation}
The average ratio $\bar r$ results from averaging  over the 
eigenstates of the Hamiltonian and over  disorder configurations. 
For Poisson-distributed energy-level spacings, e.g., for integrable systems, 
the average value of the ratio is $\bar r=2\ln(2)-1\approx0.386$. In the 
non-integrable case, one  expects that level spacings  are described 
by the Gaussian Orthogonal Ensemble (GOE) \cite{dAlessio16}. This yields $\bar r=4-2\sqrt{3}\approx0.535$ for $3\times 3 $ matrices.  

In Fig.~\ref{transition} (a), we show $\bar{r}$ \cite{Oganesyan07,Luitz14} 
as a function of $W/J$. 
The expected behavior  \cite{Oganesyan07,Luitz14} is visible. At weak disorder, $\bar r$ converges 
to the GOE result upon increasing $L$ while  in the strong-disorder 
regime, $\bar r$ is compatible with the Poisson value $\bar r\approx 0.38$. 
Using the scaling ansatz form $\bar r=g(L^{1/\nu}(W-W_c))$~\cite{Luitz14}, with $\nu$ 
a critical exponent and $W_c$ the critical value of the disorder,  
we get  $W_c/J = 2.0(2)$ (we have used $L=14,16,18$ for the scaling collapse). However, similar to 
Ref.$~$\cite{Luitz14}, one obtains $\nu=0.6(1)$, which  
violates the Harris bound \cite{Harris74,Chandran15,Khemani17}.
In conclusion, the analysis of the gap ratio $\bar r$ suggests a change in 
behavior at $W_c/J\approx 2$.

To complement our analysis, we also consider the occupation 
distances $\delta n_{i}$ and $\delta n_{\alpha}$ introduced in Ref. \cite{Hopjan20}. 
These are derived from the OPDM 
\begin{equation}
	\rho_{ij}^{(1)}=\langle\psi_n|c^\dagger_ic_j|\psi_n\rangle, 
\end{equation}
where $|\psi_n\rangle$ denotes an eigenstate. We define $n_i$ as the 
fermionic spatial occupations $n_i=\rho_{ii}^{(1)}$, and $n_\alpha$ are the eigenvalues 
of $\rho_{ij}^{(1)}$. Here, we consider the distances $\delta n_i = n_i -[n_i]$ and 
$\delta n_\alpha = n_\alpha -[n_\alpha]$ to the closest integers of $[n_i]$ and $[n_\alpha]$, respectively. 
Finally, we obtain the averaged occupation distances $\overline{\delta n_i}$ and 
$\overline{\delta n_\alpha}$ by averaging over different disorder realizations. 

The occupation distances measure the
degree  of Fock-space localization in the chosen single-particle basis. They are almost independent of system size 
in the MBL phase \cite{Hopjan20}, while in the ergodic region, $\overline{\delta n_i}$ 
must converge to the average particle filling, in our case $0.5$, 
while $\overline{\delta n_\alpha}$ approaches a smaller, energy-dependent value. 

In Figs.~\ref{transition}(b) and (c), we plot the occupation distances 
$\overline{\delta n_i}$ and $\overline{\delta n_{\alpha}}$ 
as a function of $W/J$. We observe that $\overline{\delta n_i}$ become 
almost $L$-independent for $W/J>2$. For $\overline{\delta n_{\alpha}}$,
this happens for $W/J>2.6$. Thus, the occupation distances confirm the 
qualitative scenario obtained from the analysis of the gap ratio $\bar r$ 
[see Fig.~\ref{transition}(a)].

\bibliographystyle{apsrev4-1}
\bibliography{paper_MBL_entropy}
\url

\end{document}